# Attractor structures of signaling networks: Consequences of different conformational barcode dynamics and their relations to network-based drug design

Kristóf Z. Szalay,[a] Ruth Nussinov[b,c] and Peter Csermely*[a]

**Abstract:** Conformational barcodes tag functional sites of proteins, and are decoded by interacting molecules transmitting the incoming signal. Conformational barcodes are modified by all co-occurring allosteric events induced by post-translational modifications, pathogen, drug binding, etc. We argue that fuzziness (plasticity) of conformational barcodes may be increased by disordered protein structures, by integrative plasticity of multi-phosphorylation events, by increased intracellular water content (decreased molecular crowding) and by increased action of molecular chaperones. This leads to increased plasticity of signaling and cellular networks. Increased plasticity is both substantiated by and inducing an increased noise level. Using the versatile network dynamics tool, Turbine (www.turbine.linkgroup.hu), here we show that the 10% noise level expected in cellular systems shifts a cancer-related signaling network of human cells from its proliferative attractors to its largest, apoptotic attractor representing their health-preserving response in the carcinogen containing and tumor suppressor deficient environment modeled in our study. Thus, fuzzy conformational barcodes may not only make the cellular system more plastic, and therefore more adaptable, but may also stabilize the complex system allowing better access to its largest attractor.

**1 Introduction: Conformational barcodes and network-based drug design strategies**
How are cellular pathways controlled, and their directions decided, coded and read? As an answer to this question delineating the molecular mechanisms governing these phenomena, the conformational barcode hypothesis [1] posits that all co-occurring allosteric events (including post-translational modifications, pathogen or drug binding, etc.) collectively tag protein binding or functional sites with a unique conformational barcode. This barcode is read by an interacting molecule which transmits the signal. A conformational barcode provides an intracellular address label, which selectively favors binding to one partner, and quenches binding to others, and in this way determines the pathway direction, and eventually, the cell's response and fate.

*[a] K.Z. Szalay, P. Csermely*
*Department of Medical Chemistry, Semmelweis University*
*Tuzolto str. 37-47, H-1094 Budapest, Hungary*
*\*e-mail: peter.csermely@med.semmelweis-univ.hu*
*[b] R. Nussinov*
*Basic Science Program, Leidos Biomedical Research, Inc. Cancer and Inflammation Program, National Cancer Institute*
*Frederick, MD 21702, USA*
*[c] R. Nussinov*
*Sackler Inst. of Molecular Medicine, Department of Human Genetics and Molecular Medicine*
*Sackler School of Medicine, Tel Aviv University*
*Tel Aviv 69978, Israel*

A key hallmark of cellular networks is their adaptation capacity. At the network level adaptation capacity is largely defined by the rigidity and plasticity of the network structure [2-4]. Rigid networks are unable to adapt, while plastic networks readily change their topology in response to environmental stimuli. Plastic/rigid cycles help to find the global optimum of the complex system (similarly to cooling/heating cycles of simulated annealing), and are often applied as a highly efficient adaptation mechanism of real world networks. Plastic networks dissipate incoming signals and noise very efficiently. In contrast, rigid networks transmit signals (and noise) with much less dissipation. Due to this difference, drug targeting of plastic networks (such as those of rapidly proliferating cells, like infectious bacteria or cancer) must use a "central hit" of central network nodes, since targeting peripheral nodes would dissipate the effect of the drug at the network periphery. On the other hand, drug targeting rigid networks (such as those of the differentiated cells of human tissues) at their central nodes may over-excite these networks and cause drug side-effects or even toxicity. Here a "network influence" strategy, which targets the neighbors of central nodes, may be the preferred option (see **Figure 1** for further explanation). Indeed, antibiotics often target central network nodes, while allo-network drugs (binding to neighboring nodes of the actual targets [5]) are increasingly used to cure complex diseases such as diabetes [2-5].

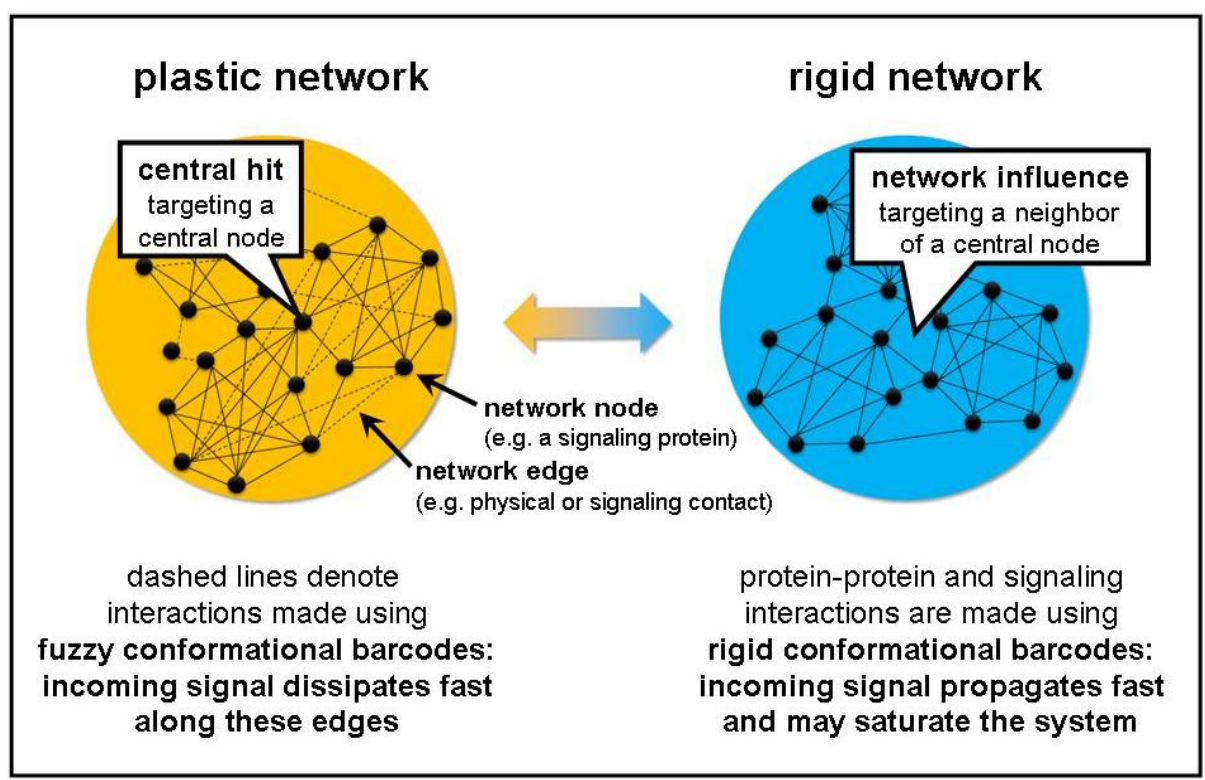

**Figure 1.** Comparison of plastic and rigid networks. The illustrative plastic network (shown on the left side of the figure) contains a number of protein-protein and/or signaling interactions, which correspond to fuzzy conformational barcodes (denoted as dashed lines in the figure). In plastic networks signals are dissipated fast preferentially along the weak interactions represented by fuzzy conformational barcodes. In plastic networks efficient drugs usually target central nodes (using the central hit strategy [3]) to allow the drug-induced signal to reach a significant segment of the network before it quickly gets dissipated. In a rigid network rigid conformational barcodes predominate (as illustrated on the right side of the figure). In rigid networks the predominantly yes/no responses of rigid protein-protein and/or signaling interactions propagate fast. Here drug targeting a central node may over-excite the network, which may induce drug side effects or toxicity. Thus, in a rigid network targeting the neighbors of central nodes by allo-network drugs [5] (using the network influence strategy [3]) is more advisable than targeting the central nodes themselves.

## 2 Classes of conformational barcode dynamics and their consequences in network-based drug design strategies

If conformational barcodes were only extremely rigid, their readout would be restricted to a simple yes/no response. In the reality of noisy cellular systems, conformational barcodes [1] represent ensembles of several conformational states [6]. Taking into account this uncertainty of the readout integrating an ensemble of conformational states, here we suggest that conformational barcodes may be rigid (ensuring a near-to-perfect readout as suggested previously [1]) but may also be fuzzy (enabling a plasticity in the final response). Rigid conformational barcodes may arise from a rigid structure of the underlying protein structure and its residue interaction network [1,2]. Fuzzy (plastic) conformational barcodes may be the consequence of structural disorder of the respective protein or protein segment [2,7]. Most conformational barcodes are in-between the above extremities of rigid or fuzzy conformational barcodes. Indeed, co-evolutionary signals across protein lineages revealed a broader range of conformational states than observed in a single protein, and showed several hidden conformations, which may have a functional importance [8,9]. Ensembles of hidden conformations extend the options of the development of fuzzy conformational barcodes beyond those given by intrinsically disordered proteins or protein segments.

Fuzzy conformational barcodes allow "yes" readout even in cases when the docking of the interacting molecule does not have 100% efficiency. The introduction of fuzzy conformational barcodes necessitates a compromise in readout accuracy and results in an increase of readout efficiency. The increased plasticity of binding recognition (characterized by a fuzzy conformational barcode) allows surpassing an otherwise too high signal detection threshold (the binding event itself and its consequences in cellular signal transduction). Such a disorder-driven extended signal sensitivity is similar to stochastic resonance (or stochastic focusing [10]), where an increased noise level helps the activation of the system even in the case of sub-threshold incoming signals.

Several factors may contribute to the rigidity/plasticity of conformational barcodes and may thus modify the plasticity of the final cellular response. Multiple posttranslational modifications (such as multi-site phosphorylation [11]), molecular crowding [12] (related to changes in intracellular water content [4]), or the intensity of molecular chaperone action (related to the occupancy of molecular chaperones by misfolded proteins [13,14]) may all modulate the fuzziness of conformational barcodes and thus the plasticity and adaptability of the final cellular response. Cellular noise may both be a reason and a consequence of increased plasticity of cellular responses [4] including conformational barcode plasticity. A recent paper [15] described the role of molecular co-chaperones of Hsp90 in the modulation of Hsp90 conformational barcode fuzziness. The co-chaperone p23 slowed down inter-domain allosteric signals of Hsp90, while another co-chaperone, Aha1 accelerated their propagation [15]. Thus, p23 rigidifies the Hsp90 conformational barcodes, and Aha1 makes them more plastic (fuzzy). Similar modifying roles were recently shown for other co-chaperones of Hsp90 regulating the recognition of Hsp90 client-proteins [16].

Rigid barcodes build rigid protein-protein interaction and signaling network structures, while fuzzy barcodes enable larger plasticity (**Figure 1**). This situation makes the two drug design strategies [3] above even more efficient. The network influence strategy applied to rigid networks needs a rather high precision, since in this strategy neighbors or second neighbors of network nodes involved in the disease are the real targets of drug action. Indirect targeting is helped by the high precision of rigid conformational barcodes characteristic of these rigid networks. Here fuzzy conformational barcodes do not compromise the precision of drug action either, since the central hit strategy targets central nodes of the networks, where a multitude of pathways converge, and allow a large variability in the transmission of drug-induced signals – irrespectively of the precision of the original target response.

**3 A case study: increased cellular noise (fuzzy conformational barcodes) has a profound effect on the population of cellular steady states (attractors)**

As a case study to show the consequences of increased plasticity of conformational barcodes at the cellular network level, we choose noise as a surrogate of fuzzy conformational barcodes, and the human signaling network as an example of cellular networks. We introduced different levels of noise to a healthy human signaling network consisting of nodes frequently involved in cancer progression [17]. In particular, we were interested in testing the robustness of the steady states (attractors) of this signaling network in an environment containing carcinogens, active growth factors, abundant nutrients, normoxia and inactive tumor suppressors (state 11100 in reference [17]). Under these conditions, a previous study [17] detected three simultaneously occurring attractors, one apoptotic and two proliferative (**Panels A, B and C of Figure 2**). (We note that in reference [17] attractors of 11100 are shown on Figure S1 as environmental condition 11101, which is most likely a typographic error, since environmental state 11101 possesses only a single apoptotic attractor, as shown on Figure 2. and Table S1 of reference [17].) Apoptosis can be considered here as the "healthy attractor", due to the carcinogenic environment.

**3.1. Simulation results**

In our analysis using our previously published Turbine network dynamics toolkit [18,19] we started the simulation of signaling network dynamics from its 10,000 randomly selected activation states, and allowed its signal transduction from these states for 200 signaling steps using the signaling model detailed in the Methods section. Using this methodology without noise we successfully replicated the previous findings [17] using a signaling model built for Turbine (details of the dynamic model are described in the Methods section). One of the two separate proliferating attractors was termed "stable" (**Panel B of Figure 2**), since only the cyclins showed cyclic behavior in that attractor, while the other proliferating attractor was termed "unstable" (**Panel C of Figure 2**), where numerous other signaling proteins were also included in the limit cycle of the attractor.

As a next step, we have tested the behavior of the three attractors by introducing low level (5 steps, ±0.01 unit), medium level (10 steps, ±0.1 unit) and high level (30 steps, ±0.4 unit) pink (1/f) noise to the system. We selected pink noise, since its scale-free distribution is close to the real nature of cellular noise [20]. Simulations were repeated 1,000 times for all noise levels and all attractors. In this way we could measure the probability of transfer between attractors with different levels of noise. At high noise levels, the transfer probabilities converged to the relative attractor basin sizes due to the high level of introduced randomness (**Panel D of Figure 2**). Measurements with a milder noise level (10 steps of added noise with ±0.1 unit) already prohibited transfer from the very stable apoptotic attractor. At these medium noise levels, the stable proliferating state had a higher chance (63.3%) of resisting the noise, and a lower, but still significant chance (36.2%) of transferring to an apoptotic orbit. However, the unstable state was still dislodged mostly towards the apoptotic attractor in all trials (**Panel E of Figure 2**). The unstable attractor was still highly unstable at the lowest tested noise level (5 steps of added noise with ±0.01 unit, **Panel F of Figure 2**) resisting the transfer only in 0.2% of the cases; at the same time, with these low noise levels, both the apoptotic and the stable proliferating attractor rejected all 1,000 transfer attempts.

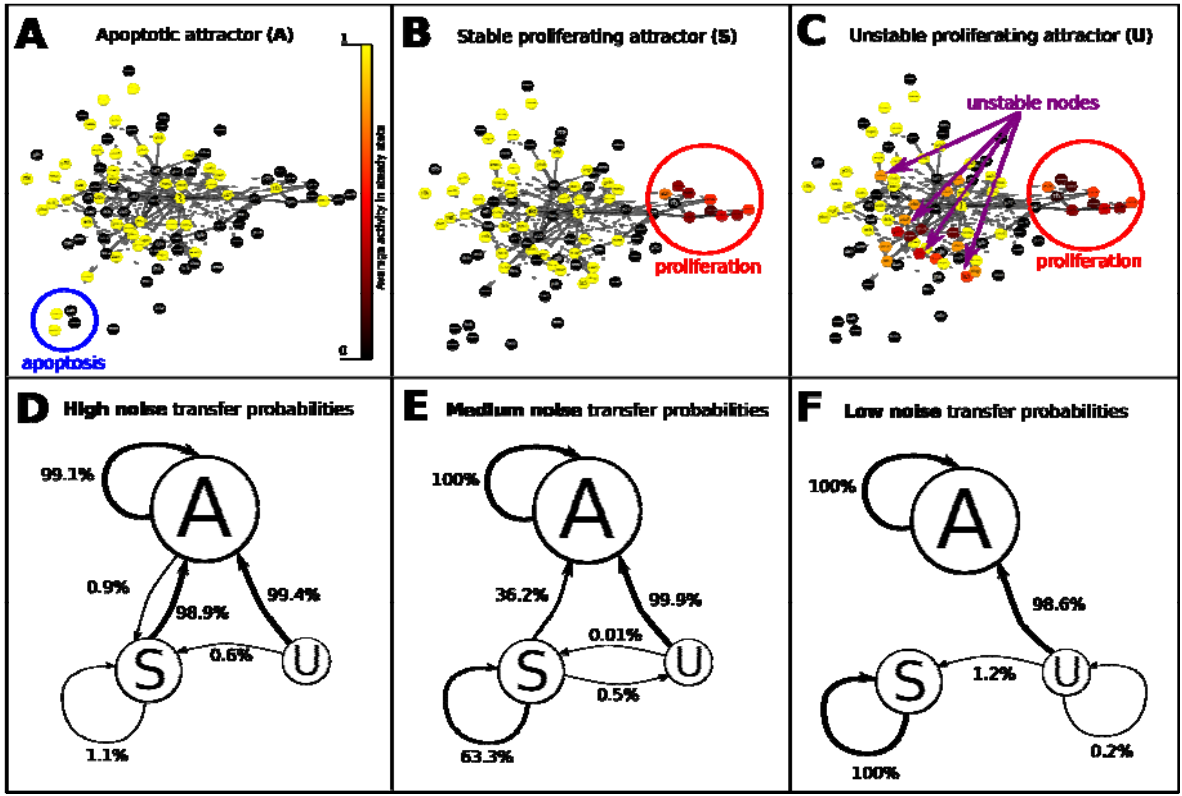

**Figure 2.** Response of the attractor structure of a healthy human cancer-related signaling network to noise. Pink (1/f) noise of three different levels (low: 5 steps, ±0.01 unit representing 1% noise level; medium: 10 steps, ±0.1 unit representing 10% noise level; high: 30 steps, ±0.4 unit representing 40% noise level) was introduced to the healthy human cancer-related signaling network described by Fumiã and Martins [17] in each of the three steady states (apoptotic, stable proliferating, unstable proliferating) 1,000 times for each steady state and noise level. The evolution of the network dynamics was observed for 200 steps, and the steady states (attractors) of the network state space were determined using our Turbine network dynamics program (www.turbine.linkgroup.hu; [18,19]). Panels A, B and C display the three steady states, Panels D, E and F show the transfer probability (the probability the steady state changes after the application of noise) for low-level, medium-level and high-level noise, respectively. The results indicate that intrinsic cellular noise levels (10 to 15%, corresponding to the medium-level noise [22]) may have a protective effect against steady states with small attractor basins, such as the proliferating attractors in this case.

### 3.2. Discussion

Theoretically, the stability of an attractor towards noise should correspond to the size of the attractor basin [21], which turned out to be true for the dominating large apoptotic attractor (relative size 0.983, measured from 10,000 random samples). Interestingly, the stable proliferating attractor not only looks more stable because fewer nodes are in its limit cycle, but also displays greater stability to noise, despite the smaller size of its basin of attraction (0.005) compared to the unstable proliferating attractor with a relative basin size of 0.012. This discrepancy can be attributed to the different noise response of the attractors. The apoptotic attractor showed a remarkable stability dissipating all introduced noise, while systems in the unstable proliferative attractor were found to amplify the introduced noise, displaying high instability. The stable proliferating attractor displayed a very interesting critical behavior at low noise levels, where the properties of the steady state remained the same after application of the noise, but the noise was not dissipated, i.e. any introduced noise still circulated in the system. Further application of outside noise could either cancel or (in most cases) reinforce the remaining noise, eventually building up the noise to levels high enough to cause the system to transfer to the apoptotic attractor, dissipating all noise. In a strict sense this results in infinite number of stable proliferative attractors, leading us to simplify all these attractors under a single stable proliferating attractor. Noisy stable proliferating attractors did not appear in the random sampling of states during the measurement of the attractor basin size, since we have only used full activation (1) and no activation (0) without intermediate values in these initial measurements carried out for easier testing and to be more comparable with the previous article [17].

**3.3. Conclusions of the case study**
Altogether, our tests indicate that intrinsic cellular noise (estimated to be approximately 10 to 15% [22] corresponding to medium level of noise in our experimental setup) may have a protective, optimizing effect on healthy cells, enabling the signaling network (and possibly other cellular networks too) to cross threshold barriers of diseased attractors and land in the largest attractor of the healthy cell (the healthy attractor in our case is the apoptotic one due to the carcinogenic environment). Thus, fuzzy conformational barcodes may not only make the cellular system more plastic, and therefore more adaptable, but (if present at intermediate levels) may also stabilize the complex system allowing a better access to its largest attractor. These are good news, since one would expect Darwinian evolution to shut down "abused" cells, irrespective of the nature of the perturbation. The fact that our simulations comfort this common-sense expectation is an encouraging proof that, in spite of the many simplifications in the working hypotheses, the dynamic model, and the absence of the complete set of quantitative kinetic parameters, the network dynamics model used had enough information content (due to its topology and related basic dynamics) to mimic some key aspects of living systems.

**4 Methods**
The trajectories of the network were simulated using the Turbine toolkit (http://turbine.linkgroup.hu [18,19]) using *Equation 1* as the update rule. The coefficients of the equations were transformed to edge weights having the same value as the original coefficient. A helper node named "1", constitutively active in all experiments, was introduced for setting the activation threshold values. The signaling network was constructed using the network of Fumiã and Martins [17], and contained 97 nodes and 295 directed, weighted edges. The additional node represented a "threshold-node" with 46 new edges leading to nodes of the original network having a non-zero threshold value. This model was implemented to make these results compatible with those obtained using other network dynamic models such as SignaLink 2.0 [23] or a T-cell leukemia network [24], which are systems containing no signaling thresholds. The system was evolved according to *Equation 1* (where $A$ is the activity level of the updated node, $l$ is the in-degree of the observed node, $\omega$ is the edge weight of the current link and $A_l$ is the parent node of the current inbound link) with a maximum activity level of 1, a minimum activity level of 0, and a deactivation rate ($D$) of 1 (complete deactivation in the step following the cessation of the last activation). The use of *Equation 1* results in the same behavior as the update equation in article [17].

$$\frac{dA}{dt} = \sum_l \omega A_l - AD \qquad (Equation\ 1)$$

The size of the basins of attraction were calculated by evolving the network for 200 time steps from 10,000 random states, where each node could have an activation level of 0(no activation) or 1(full activation), but no intermediate values. Maximum detection length for limit cycles was set to 40 (at least two full periods observed in this time window), point attractors were tested for steadiness for 20 time steps. Pink noise was created with the noise command of the tuneR package [25] of the R toolkit [26], using only the left channel of the generated noise multiplied by the requested noise amplitude, and a different realization of the noise for each node in every simulation.

**Acknowledgments**
Authors thank to the anonymous reviewers of this paper for their suggestions, and Reviewer 1 highlighting an important conclusion of the current work on the robustness of the Turbine method, which we inserted as the concluding two sentences of our paper. Daniel V. Veres's (Department of Medical Chemistry, Semmelweis University, Budapest, Hungary, www.linkgroup.hu) contribution to Figure 1 is kindly acknowledged. This project has been funded in whole or in part with Federal funds from the National Cancer Institute, National Institutes of Health, under contract number HHSN261200800001E and by a research grant from the Hungarian National Science Foundation (OTKA K83314). The content of this publication does not necessarily reflect the views or policies of the Department of Health and Human Services, nor does mention of trade names, commercial products, or organizations imply endorsement by the U.S. Government. This research was supported (in part) by the Intramural Research Program of the NIH, National Cancer Institute, Center for Cancer Research.

**Conflict of interest**
K.Z.S. and P.C. declare that a patent application (reference [19]) has been submitted on the Turbine network dynamics toolkit used in this paper. However, the current use of Turbine is not directly related to its applications covered by the patent.